\documentclass[twocolumn]{openjournal}
\usepackage{natbib}
\usepackage{graphicx,amsmath,amssymb,amstext}
\usepackage{amsbsy,amsfonts,amsthm,color}
\usepackage[colorlinks,linkcolor=blue,citecolor=blue,urlcolor=blue ]{hyperref}
\usepackage[utf8]{inputenc}
\usepackage{float}
\usepackage{comment}
\usepackage{xcolor}
\usepackage{ulem}
\usepackage[T1]{fontenc}
\usepackage[caption=false]{subfig}

\newcommand{\msun}{\ \text{M}_\odot}

\newcommand{\msolar} {$\rm{M_{\odot}}~$}
\newcommand{\msolarc} {$\rm{M_{\odot}}$}

\newcommand{\arepo}{\texttt{Arepo~}}
\newcommand{\enzo}{\texttt{Enzo~}}
\newcommand{\renaissance} {\texttt{Renaissance~}}

\begin{document}


\title{Growth of Light-Seed Black Holes in Gas-rich Galaxies at High Redshift\vspace{-3em}}

\author{Daxal Mehta$^{* \ 1,2}$}
\author{John A. Regan$^{1, 2}$}
\author{Lewis Prole$^{1, 2}$}
\email{$^*$email: daxal.mehta.2024@mumail.ie}

\affiliation{Department of Physics, Maynooth University, Maynooth, Ireland}
\affiliation{Centre for Astrophysics and Space Science Maynooth, Maynooth University, Maynooth, Ireland}


\begin{abstract}
\noindent Recent observations by the James Webb Space Telescope confirm the existence of massive black holes ($>10^6$ $\msun$) beyond the redshift of $z=10$. However, their formation mechanism(s) still remains an open question. Light seed black holes are one such formation pathway, forming as the end stage of metal-free (Population III) stars. Light seed black holes can grow into massive black holes as long as they accrete near the Eddington limit for substantial periods or undergo several bursts of super-Eddington accretion. In this work, our aim is to ascertain if light seeds can grow in gas-rich galaxies - similar to those expected at high redshift (z $\gtrsim 10$). Using the \arepo code, we follow self-consistently the formation of Population III stars and black holes in galaxies with total masses in the range $10^8$ \msolarc.  We find that in the absence of feedback, black holes can grow to $10^5$ $\msun$ in just $10^4$ yr. These black holes do not decouple from the gas clumps in which they are born and are able to accrete at hyper-Eddington rates. In the presence of supernova feedback, the number of actively growing black holes diminishes by an order of magnitude. However, we still observe hyper-Eddington accretion in approximately 1\% of the black hole population despite supernova feedback. This (idealised) work lays the foundation for future works, where we will test our models in a cosmological framework. In this work, we neglect radiative feedback processes from stellar evolution and from accretion onto the growing black holes. This likely means that our results represent an upper limit to light seed growth. We will address these shortcomings in future work.
\end{abstract}


\maketitle

\section{Introduction} 
\label{sec:intro}
\noindent Quasars at  redshifts, $z \sim$ $6-7$ have been observed to host (super-)massive black holes (MBHs)\footnote{We use the term MBH here to refer to central or near-central black holes with masses in excess of $10^3$ \msolarc} of mass $10^{8-10}$ $\msun$ \citep{banados2018800, farina2022x, mazzucchelli2023xqr, wang2021revealing, yang2020poniua, yang2023spectroscopic} (see also review by \citealt{fan2023quasars}). Moreover, recent JWST observations at redshift of $z\sim8-11$ have unearthed galaxies with central MBHs with masses of $\gtrsim 10^6$ $\msun$ \citep{maiolino2024small, Larson_2023,Goulding_2023, Bogdan_2024}. These observations imply that MBHs must have grown rapidly from their seed masses at very early epochs. The physical origin of the seeds is still uncertain with competing models including remnants of Population III (PopIII) stars (Light Seeds - see e.g. \citealt{madau2001massive}), run-away mergers of stellar bodies in dense star clusters \citep[e.g.][]{devecchi2009formation}, or direct collapse of supermassive stars of $10^4-10^6$ $\msun$ (Heavy Seeds, see e.g. \citealt{haehnelt1993formation, ferrara2014initial, mayer2018route, woods2017maximum, Woods_2019}). For this paper, we define light seeds as those with (initial) masses of less than 1000 \msolar and heavy seeds as those with masses in excess of 1000 \msolarc.

Lights seeds are expected to be formed at early times, starting anywhere from as high a redshift  as $z \sim 30$ \citep{haiman2001highest, madau2001massive, heger2003massive, volonteri2003assembly, madau2004early, Volonteri_2012,schauer2021influence} in low-mass dark matter halos of $\gtrsim 10^6$ $\msun$. These early halos will eventually merge to become some of the most massive halos in existence today. Their progenitor (Pop III) stars exhibit mass ranges from tens to hundreds of solar masses \citep{bromm1999forming, abel2000formation, abel2002formation, bromm2002formation, o2007population, turk2009formation, clark2011formation, hirano2014one,prole2023dark}. Light seeds form through a multitude of pathways owing to the wide range of Pop III stellar masses including core-collapse supernovae (SNe) or hypernovae ($11 \msun \leq M \leq 40 \msun$, \citealt{woosley1995evolution, nomoto2006nucleosynthesis}) and direct collapse into stellar mass black holes (BH)\footnote{We use the abbreviation BH to refer implicitly to stellar mass black holes or light seeds} ($40 \msun \leq M \leq 140 \msun$ and $M > 260 \msun$, \citealt{heger2002nucleosynthetic}). These seeds can grow into a MBH if they have sustained near-Eddington accretion on Gyr time periods. However, this condition is non-trivial to satisfy in dark matter halos as feedback from progenitor stars, SNe explosion, BH accretion and BH dynamics affects the supply of gas to the BHs \citep{wise2008resolving, alvarez2009accretion, milosavljevic2009accretion, smith2018growth}, resulting in highly sub-Eddington accretion for long periods. The growth can be accelerated if the BHs can accrete at hyper-Eddington rates ($\rm{f_{Edd}} \gtrsim 1000$, see theoretical arguments by \citealt{inayoshi2016hyper, jiang2019super, park2020biconical, kitaki2021origins, botella2022structure}). However, these theoretical studies assume that the BH is able to pull gas from scales larger than their accretion disk, or equivalently that gas flows rapidly onto the BHs from large distances. It is not clear if the BH will encounter dense regions of gas for such high accretion rates to be realised. If the BH does encounter such regions, it can grow into a MBH on timescales of less than 500 Myr \citep{Lupi_2016}, or perhaps even less due to a rapid runaway accretion episode \citep{shi2023hyper}. \\

\indent More exotic, or equivalently rarer, channels through which MBH seeds can be formed can avoid the need for super- or hyper-Eddington accretion by forming with initially much heavier masses - so-called heavy seeds \citep{regan2024massive}. As defined above, heavy seeds are seeds with masses in excess of 1000 \msolarc. These can form from a number of potential channels, including runaway collapse inside a dense stellar cluster \citep{PortegiesZwart_2004, devecchi2009formation, katz2015seeding, Rizzuto_2021, Arca-Sedda_2021, Gonzalez_2021, Reinoso_2023, Arca-Sedda_2023, Rantala_2024} or through runaway collisions inside a BH cluster \citep{Fragione_2018, Fragione_2022, Antonini_2019, Mapelli_2021, Arca-Sedda_2021}. Both of these avenues have been extensively modelled over the previous two decades and while a full consensus is still lacking, the formation of MBH with masses in excess of $\sim 1000$ \msolar appears viable.  \\
\indent Another route to heavy seeds is via the intermediate stage of a very massive or supermassive PopIII star \citep{regan2009formation, regan2009pathways,regan2014numerical, spaans2006pregalactic, latif2013high,latif2013black, becerra2015formation, Chon_2016, Chon_2017, Chon_2017b, Chon_2020, Woods_2018, woods2017maximum,Woods_2019, Wise_2019, Regan_2020b}. In this case a star with a mass between $10^{3-5}$ \msolar forms and collapses directly to a near-equivalent mass black hole. Such heavy seeds are regularly invoked to explain the existence of MBH at high-z, while the chemical signatures of the progenitor super massive star (SMS) phase may already have been detected \citep{Nandal_2024, Nandal_2024b, Schaerer_2024}. However, in this study, we will focus on light seeds and how rapid growth onto light seeds can provide an additional channel to MBH seeding. 

To understand the potential hurdles to the growth of light seeds, we study here whether rapid growth of light seeds can be achieved within gas-rich galaxies at high redshift. We conducted simulations using the \arepo code \citep{Springel_2010, Pakmor_2016}, incorporating self-consistent PopIII star formation, MBH formation, as well as SNe feedback. Our newly developed modules are designed to simulate the creation and accretion of sink particles, which serve as analogues to stars during their stellar lifetimes and subsequently evolve into BHs.

The initial conditions for our simulation involve a galaxy with a mass of approximately $10^8 \msun$, generated using the GalIC tool \citep{yurin2014iterative}, which enables the construction of stable, self-consistent galaxy models. We performed two distinct realisations of this galaxy to explore the impact of stellar feedback mechanisms. In the first scenario, SNe feedback is disabled, allowing us to observe the growth of BHs in the most ideal conditions. In the second scenario, SNe feedback is enabled, providing insight into how such feedback influences the conditions conducive to light seed formation and growth.

The paper is structured as follows. In \S \ref{sec:method}, we discuss the numerical methodology of our work. We briefly describe the working of the \arepo code. We also discuss in depth the initial conditions used and the sink particle techniques employed. \S \ref{sec:results} describes our results for two realisations (with and without SNe feedback). We compare our study with others from the literature in \S \ref{sec:discussion} and discuss some of our caveats in \S \ref{sec:caveats}. Finally, we conclude our work in \S \ref{sec:conclusion}.

\section{Numerical methods}

\label{sec:method}
To carry out the simulations described in this paper, we use the moving mesh code \arepo \citep{Springel_2010, Pakmor_2016}, which solves the Euler equations on an
unstructured Voronoi tessellation that moves with the flow. The fluid dynamics are solved using a finite-volume, second-order reconstruction scheme coupled with an exact Riemann solver. The unstructured mesh adapts by locally refining and de-refining in a quasi-Lagrangian way to maintain a typical gas mass target per cell. We employ the hierarchical time integration scheme described in \cite{Springel_2021}, which splits the Hamiltonian into “fast” and “slow” components, i.e. those on short versus long time-steps, respectively. With this approach, interactions between particles/cells on the smallest time steps (up to a maximum of 3000 particles) are solved using direct summation, while a standard octree approach is used for all other interactions. This is primarily done for efficiency, to avoid unnecessary tree constructions for small numbers of particles, but has the added benefit of more accurately calculating the gravitational forces between black holes and their nearest neighbours, which are on the shortest time-steps, via direct summation \citep{bourne2024dynamics}. \\

\indent Within our suite of simulations, the dark matter (DM) particles have physical gravitational softening lengths $\Delta x$ = $10^{-2}$ pc, while gas cell softening adapt with the cell size down to a minimum of $\Delta x$ = $10^{-2}$ pc. This is the minimum cell length we allow and is controlled by the \texttt{MinVolume} parameter. Refinement is controlled via the \texttt{JeansRefinement} parameter with the number of cells used to resolve the Jeans length of the gas set to 8. We also employ a gas temperature floor of $\sim 3$ K.

\subsection{Initial conditions} \label{sec:ics}
\noindent We create the initial conditions (ICs) for our idealised and isolated galaxies using GalIC \citep{yurin2014iterative}, which constructs galaxy models in collisionless equilibrium made up of both DM and other collisionless particles which make up the disk and bulge components. To accurately and realistically model early high redshift galaxies, we initialise our galaxies using the results from \citealt{chen2014scaling}, who found that for halo masses between $10^7$ \msolar and $10^{8.5}$ \msolar, the gas fraction to the total halo mass is $0.1$, lower than the cosmic baryon fraction. The reason for this is that in small early galaxies, SNe feedback is very effective in removing gas from galaxies, at least, until galaxies become more massive. As a result, we create our idealised galaxies with $90\%$ of the galaxy's mass in the DM halo. The remaining mass of the galaxy being baryonic. 

The DM halo is modelled as a NFW \citep{NFW} halo profile, while the gas is modelled as a spherical bulge with a radius $0.1$ times the radius of the parent halo. We set the virialized circular velocity of the halo to be $10$ km/s, which constrains the total mass of the galaxy to be $2.32 \times 10^8$ $\msun$. 
We create the IC using $10^6$ DM particles and $10^4$ gas particles.  The (initial) mass resolution is then $\sim 200$ \msolar and $\sim 2000$ $\msun$ for the DM and gas particles, respectively. During the course of the simulation, the mass resolution of the gas cells refine based on local gas properties (specifically the Jeans length) and more gas cells are added to the grid as discussed in \S \ref{sec:method}. The refinement causes the gas cell mass to range from approximately one to a few thousand solar masses. The DM particle resolution remains fixed throughout the simulation, but at $\sim 200$ \msolarc, it is highly resolved. In addition to the DM particles and gas particles we also include $10^4$ collisionless `star' particles which reside in a plane around the halo centre. These "stars" model the previous generation of stars inherited via the hierarchical build up of the galactic structure. The total mass in these genetic stars is $2.32\times10^6$ and they add extra potential to the halo centre, helping to stabilise the inner part of the galaxy. The different components of the halo are outlined in Table \ref{tab:ICs}.

\begin{table}
    \centering
    \begin{tabular}{|p{1.5cm}|p{2.2cm}|p{1.2cm}|p{1.5cm}|}
    \hline
       Component  & Total Mass [\msolarc] & Particle Number & Particle Mass [\msolarc] \\
       \hline
        DM & $2.09\times10^8$ & $10^6$ & $209.21$ \\
        Gas & $2.09\times10^7$ & $10^4$ & $2092.11$ \\
        Stars & $2.32\times10^6$ & $10^4$ & $232.46$ \\
        \hline
    \end{tabular}
    \caption{Particle data for the initial condition galaxy created using GalIC.}
    \label{tab:ICs}
\end{table}

Following the creation of the initial conditions using GalIC, the output must be re-formatted into an output suitable for \arepo. To enable this, GalIC generated ICs must be inserted into our 20 kpc box and evolved using \arepo, this requires that the smoothed particle hydrodynamic (SPH) particles created by GalIC be converted into mesh cells. This is achieved using the \texttt{ADD\_BACKGROUND} functionality \citep{Springel_2010, weinberger2020arepo} in \arepo. This procedure of taking the idealised halo and inserting it into an \arepo box inevitably results in some mass loss from the halo. This is because a small number of particles are scattered far from the host halo due to dynamical interactions at the outskirts of the halo. Therefore, these particles leave the 20 kpc box (isolated boundary conditions) in which we conduct the \arepo simulation. The final result is therefore a halo of mass $1.78\times 10^8$ $\msun$ in a box of $20$ kpc.

\subsection{Isolated Galaxy Evolution}
Once the initial conditions have been created and read in by \arepo, the galaxy is evolved initially for $100$ million yr at a resolution of $1$ $\rm{pc}$ without the sink particle routines turned on. This was done to ensure that the galaxy has a suitably chaotic nature, similar to what is observed for high-z galaxies. Within this period, the spherical bulge of gas collapses to form a turbulent planar structure. The state of the galaxies at this time can be seen in Figure \ref{fig:gas_projection}. Once the simulations have evolved for 100 Myr, we then restart the simulation at our canonical resolution of $10^{-2}$ $\rm{pc}$ cell size with the sink particle routines turned on (see \S \ref{Sec:sink_particles}). We perform a total of two separate realisations- one with SNe feedback turned off and the other with SNe feedback turned on. The high resolution ($10^{-2}$ $\rm{pc}$) of our simulation prohibits us from running the simulations for extended periods, so we evolve the simulations for only 10 Myr. However, this is justified as the free-fall time of our halo is approximately 5 Myr, and we also do not account for cosmological (major) mergers, which would be expected to occur on a timescale of a few times the dynamical time of the halo.

\subsection{Sink Particle Technique} \label{Sec:sink_particles}
\noindent The sink particle technique employed in this work builds and improves on the scheme originally introduced into \arepo by \cite{wollenberg2020formation} and \cite{tress2020simulations}. Sink particles are formed in high density gas regions when a gas cell can no longer refine upon reaching the minimum cell size of the mesh. They are point like particles artificially introduced to model the behaviour of physical processes that occur below the resolution of our mesh. The original scheme introduced into \arepo converts a cell directly into a sink particle once three conditions are met: 1) the cell reaches a user defined threshold density 2) the cell is sufficiently far away from a pre-existing sink so that their accretion radii do not overlap, and 3) the gas within this region is gravitationally bound and collapsing. The approach taken here is broadly similar with the criteria extended somewhat based on the implementation of \cite{krumholz2004embedding} and previously introduced into the \enzo code \citep{Enzo2014, Enzo2019} by \cite{regan2018rise}. In this case, we change the formation criteria so that a cell is converted into a sink only after the Jeans density threshold is cleared. This removes the user-defined threshold and in doing so makes the sink formation more self-consistent. The new formation criteria checks for conditions 2, 3 as mentioned above along with these additional conditions:

\begin{enumerate}

    \item We first check that the candidate cell is at the highest refinement level and can no longer be refined. 
    \item The density of the gas cell should be higher than the local Jeans density defined by the criterion \citep{truelove1997jeans},
        \begin{equation}
            \rho < \rho_J = J^2 \frac{\pi c_s^2}{G\Delta x^2},
            \label{eq:Jeans_density}
        \end{equation}
        where $\Delta x$ is the cell size and $J=0.25$ is a constant \citep{truelove1997jeans}.
\end{enumerate}

This technique, as noted above, follows the formation technique introduced in \cite{regan2018rise} (dubbed SmartStars). In addition to the augmented formation criteria, the type of sink particle formed can be discriminated into four different flavours, namely: population II (PopII), population III (PopIII), super-massive star (SMS), and massive black hole (MBH). For the purposes of this study, we only allow (metal-free) PopIII star formation followed by the transition of the sink into a MBH particle where appropriate (see \S \ref{sec:supernova}). Henceforth, whenever a sink particle is in its stellar phase, we call it a PopIII particle, and after the stellar lifetime has ended, we call it a MBH particle.

\subsection{Assigning Sink Particle Masses}
Upon the formation of a PopIII particle, the mass of the corresponding host cell is directly transferred to the newly created PopIII particle, effectively converting the host cell into the particle itself, as outlined in the preceding discussion. This is the dynamical mass of the PopIII particle. However, we need to be careful here. Our spatial resolution ($\Delta x = 10^{-2}$ $\rm{pc}$), while good, is far from the scale of actual star formation. As a result, we additionally assign stellar masses to each PopIII particle. The stellar masses are randomly sampled from a top-heavy initial mass function (IMF). 
\begin{equation}
    \rm{f(log M) dM = M^{-1.3} exp \Big[ \big( \frac{M_{char}}{M} \big)^{1.6} \Big] dM}
\end{equation}

To construct the IMF, we set the upper and lower limits at $300\msun$ and $1\msun$, respectively, and its slope is $-1.3$. We set the characteristic mass, $\rm{M_{char}}$, to 20 \msolarc. We choose a lower limit of $1\msun$ due to the constraint of cell size resolution. At a maximum resolution of $10^{-2}$ pc, we are unable to resolve structures significantly below $1\msun$ and hence we set this as the lower limit for star formation in cells. Each PopIII particle is also assigned a stellar life according to its stellar mass. The stellar lifetimes of PopIII stars range from a few times $10^6$ to a few times $10^7$ yr, which is of the order of our entire simulation. To account for this and allow BH remnants to form sufficiently rapidly, we divide the stellar lifetimes by 100. This modification can also be thought of to account for the previous generation of star formation missed through our idealised setup. We also do not account for radiative feedback from the PopIII stars in this work (although we do account for supernova feedback). This is a limitation of our work and is discussed in detail in \S \ref{sec:caveats}

At the end of its lifetime, each PopIII particle must transition to either a MBH particle or be deleted following a Pair Instability SNe (PISN) event depending on whether or not feedback is turned on. In the case of no-feedback simulation, all PopIII particles are converted to MBH particles including those that would potentially undergo PISN. If feedback is turned on, the PopIII particle may experience a SNe (depending on its stellar mass) and inject energy into the surrounding gas (see \S\ref{sec:supernova} for more information). If a PopIII particle is in the stellar mass range where it should transition to a MBH then the MBH can accrete the surrounding gas following its formation. Only MBH particles can accrete. Mergers between MBHs are also possible via our framework, but in this study we choose not to implement MBH mergers and instead treat them as collisionless particles. We leave a study of MBH mergers in high-z galaxies to a follow-up work. 

\begin{figure*}
    \centering
    \includegraphics[width=\textwidth]{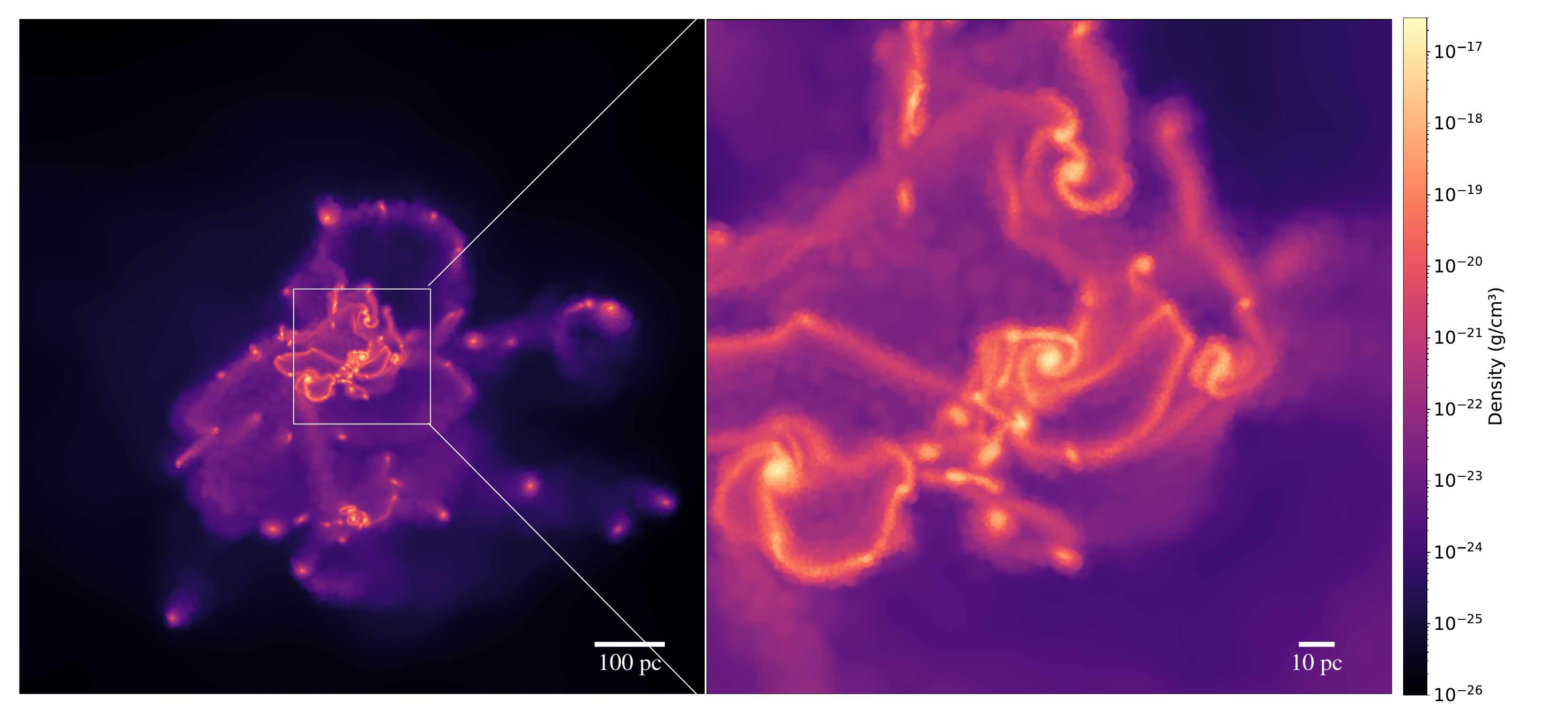}
    \caption{Left hand panel: Face-on projection of the parent galaxy that we use in both of our simulations (feedback and no-feedback). The projection is created using the density weighing scheme. This is the state of the galaxy immediately before we evolve it at our highest resolution ($\Delta x = 10^{-2}$ pc). Sink formation is not activated at this point. Right hand panel: The same image but zoomed in to the densest gas clump of the galaxy.}
    \label{fig:gas_projection}
\end{figure*}

\begin{figure*}
    \centering
    \includegraphics[width=\textwidth]{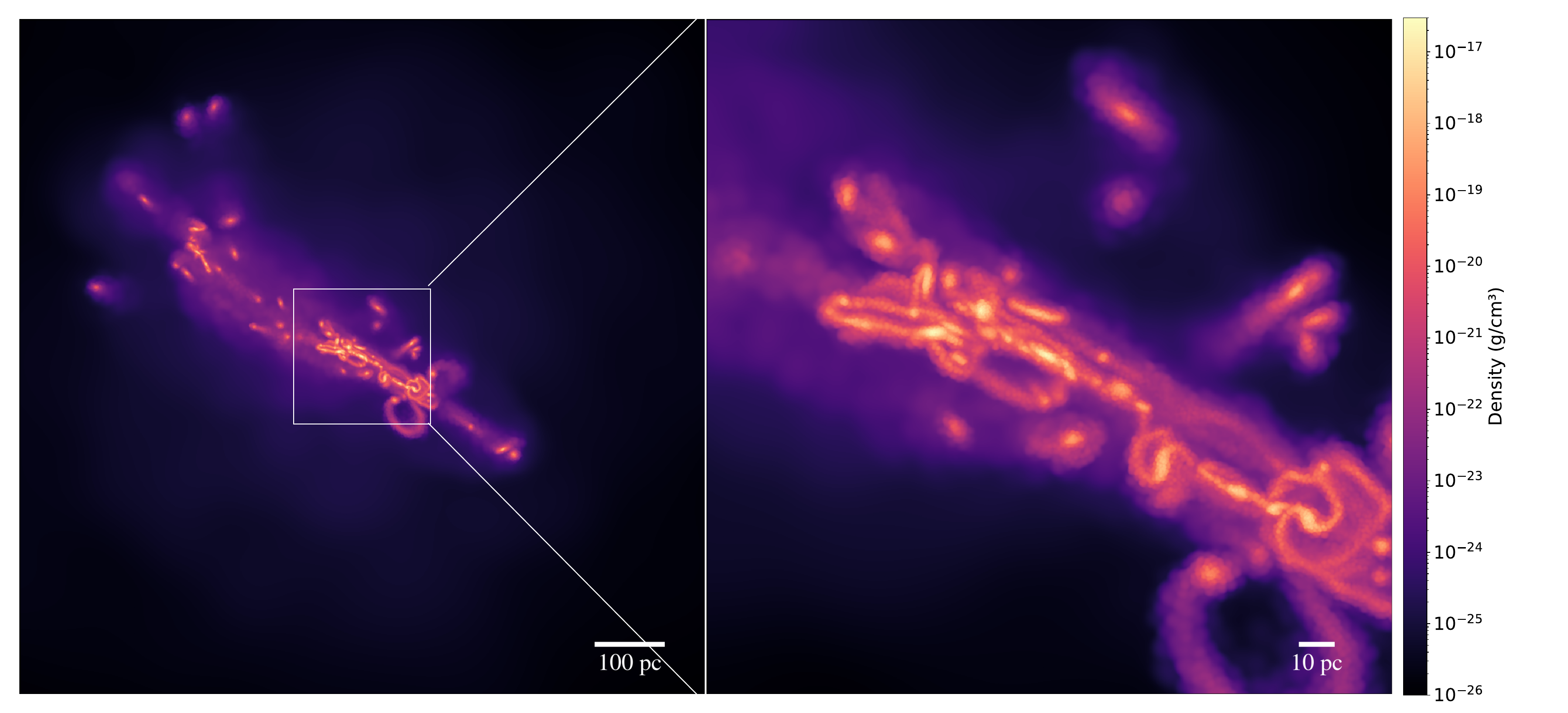}
    \caption{Left hand panel: Edge-on projection of the parent galaxy that we use in both of our simulations (feedback and no-feedback). This visualisation is identical to Figure \ref{fig:gas_projection} apart from the orientation.} 
    \label{fig:gas_projection_edge}
\end{figure*}

\subsection{Supernova Feedback}
\label{sec:supernova}
After the stellar lifetime assigned to a PopIII particle has passed, it may undergo a SNe explosion depending on its stellar mass. Following a SNe explosion, the PopIII particle either transitions into a MBH particle or is deleted (following a PISN event).
The SNe categories and energy releases are as follows:
   \begin{enumerate}
      \item PopIII particles in the stellar mass range 11-20 $\msun$ undergo a type II SNe, with energy $E_{SN}$=$E_{51}$=$10^{51} $ ergs. They transition into a MBH particle following the explosion. The MBH particle mass remaining after the explosion is the helium core mass, calculated as \citep{nomoto2006nucleosynthesis}
      \begin{equation}
    m_{\rm He} = 0.1077 +  0.3383 (m_{\star}-11),
    \label{eq:He_core_small}
    \end{equation} 
    where $m_{\star}$ is the stellar mass. 
    \item PopIII particles in the mass range of 20-40 $\msun$ undergo a hypernova. We extrapolate the values of $E_{\rm{SN}}$ and m$_{\rm He}$ from \citealt{nomoto2006nucleosynthesis}. PopIII particles in this range transition to a MBH following the explosion. The MBH mass is set to the helium core mass post explosion.
    \item PopIII particles in the mass range of $40-140$ $\msun$ experience a direct collapse into a BH and exhibit no SNe explosion. The total mass of the PopIII particle is converted into the MBH particle.
    \item Between 140 - 260 $\msun$, PopIII particles undergo a pair-instability SNe, with their values of $E_{\rm{SN}}$ and m$_{\rm He}$ calculated as
      \begin{equation}
      m_{\rm He} = \frac{13}{24} (m_{\star} - 20)
      \label{eq:He_core_large}
      \end{equation}
      \begin{equation}
      E_{\rm SN} = 5 + 1.304 (m_{\rm He} - 64).
      \label{eq:SN_E51}
      \end{equation}
    These particles are not converted to MBHs after the explosion and are instead deleted from the simulation.
    \item Finally, PopIII particles with masses above 260 $\msun$ also fall into the direct collapse category and are converted into a MBH particle without a SNe explosion.
    \end{enumerate}
Unlike previous works that ensure the Sedov-Taylor scale is resolved so that the energy of the SNe can be directly injected into the simulation mesh (see, e.g. \citealt{gatto2015modelling, tress2020simulations, magg2022metal}), we instead opt to model the explosion as an injection of momentum. As in \cite{magg2022metal}, we inject the momentum into the nearest 1000 cells surrounding the PopIII particle. At our resolution, this corresponds to an injection region of $\sim$ 0.1 pc. We calculate the momentum of the blast according to the SNe energy, the density and the metallicity of the injection region as \citep{smith2019cosmological}. 
\begin{equation}
    p_{\rm SN} = 3.0 \times 10^5 E_{\rm 51}^{16/17} n_{\rm SN}^{-2/17} Z_{\rm SN}^{-0.14} \msun \rm{km/s},
    \label{eq:mom_inj}
\end{equation}
where $E_{\rm 51} = E_{\rm SN}/10^{51}$, $n_{\rm SN}$ is the number density of the injection region, and $Z_{\rm SN}$ is the maximum between $Z/Z_\odot$ and $0.01$. $Z$ is the metallicity of the injection region, which is $0$ as we do not have metals in the simulations. As a result, we also choose a value of $Z_{SN} = 0.01$. Finally, we fully ionise the gas within the injection region and set the temperature to $10^4$ K. We then update the total energy of the cells to reflect the new kinetic and internal energies of the gas. In this study, no injection of metals or gas into the inter stellar medium is accounted for and is instead left for a forthcoming study.

\subsection{Sink Particle Accretion}
Once a PopIII particle has turned into a MBH particle, we accrete onto it using the Bondi-Hoyle-Lyttleton (BHL) \citep{bondi1952spherically} method following the prescription given in \cite{krumholz2004embedding}. Accretion onto a MBH particle then occurs by removing gas from the surrounding accretion sphere. The size of the accretion sphere is 4 cells wide. 

The accretion rate onto the MBH particle can then be calculated using the usual Bondi formula
\begin{equation}
    \Dot{M}_{Bondi} = 4 \pi \rho_{\infty} r_{BH}^2 ((1.12 c_{\infty})^2 + v_{\infty}^2)^{1/2}
    \label{eq:bondi-hoyle-acc}
\end{equation}
where $\rho_{\infty}$ is the weighted density inside the accretion sphere, $r_{BH}=\frac{G M_{sink}}{v_{\infty}^2+c_{\infty}^2}$ is the BHL radius,  $v_{\infty}$ is the mass weighted velocity and $c_{\infty}$ is the sound speed of the surrounding gas. For $v_{\infty}$, we take the relative velocity of the MBH particle and the host cell, and for $c_{\infty}$, we take the local sound speed. To calculate the value of $\rho_{\infty}$, we need to be more careful, and here we follow the procedure outlined in \cite{krumholz2004embedding}. The value of $\rho_{\infty}$ is calculated by assigning weights to all cells within the kernel radius, $r_K$. The kernel radius is given by: 
\begin{equation}
\rm{r_K} = \left\{ \begin{array}{lcr}
  \Delta x/4 & &r_{BH} <  \Delta x/4\\
  r_{BH}  & & \ \ \ \ \Delta x/4 \le r_{BH} \le r_{acc}/2\\
  r_{acc}/2 && r_{BH} > r_{acc}/2
\end{array} \right.
\end{equation}
    
For all but the very smallest black hole masses (100 \msolar MBH $\sim 0.1$ pc BHL radius), we can resolve the BHL radius with our resolution. Once the kernel radius is determined, we use this to assign a weight to every cell within our accretion sphere using the formula:
\begin{equation}
    \omega \propto \rm{exp(-r^2/r_{K}^2)}
\end{equation}
where $r$ is the distance from the cell, within the accretion sphere, to the MBH particle. The proportionality symbol is used here to indicate that the actual weight assigned is relative to the other cells in the accretion sphere. The value of  $\rho_{\infty}$ is then assigned by applying the following formula:
\begin{equation}
     \rho_{\infty} = \bar{\rho} * \omega
\end{equation}
where $\bar{\rho}$ is the mass weighted mean density within the accretion sphere. This value of $\rho_{\infty}$ is used in calculating the accretion rate (Eqn \ref{eq:bondi-hoyle-acc}) at each timestep.  In order to improve our scheme further, we also include the impact of vorticity on our accretion rate. 

\subsection{Vorticity}
Given the somewhat chaotic nature of high-z galaxies, we also take into account the vorticity, $\omega$, of the surrounding gas. To determine the vorticity of the gas, we calculate the mass-weighted curl of all gas cells following the prescription set out in \cite{Krumholz_2006}. The vorticity, $\omega$, is defined as 
\begin{equation}
    \omega = | \nabla \times \Vec{v} |
\end{equation}
with the dimensionless vorticity $\omega_*$ given by
\begin{equation}
\omega_* =  \omega * \frac{r_{BH}}{c_{\infty}}.
\end{equation}

Following \cite{Krumholz_2006} we introduce a damping factor f($\omega$) defined as,
\begin{equation}
    f_{w} = \frac{1}{1 + \omega_*^{0.9}}
\end{equation}
and calculate the accretion rate in a turbulent medium according to,
\begin{equation}
    \Dot{M}_{\omega} = 4  \pi  \rho_{\infty}  r_{BH}^2  c_{\infty} * 0.34  f_{\omega_*}
\end{equation}
Finally, the total accretion rate onto the MBH particle is,
\begin{equation}
    \Dot{M} = (\Dot{M}_{Bondi}^{-2} + \Dot{M}_{\omega}^{-2})^{-0.5}
\end{equation}

For a given time step, $t_h$, the mass of the MBH particle increases by $t_h * \Dot{M}$.  For each gas cell, we calculate the fraction of mass lost to the MBH particle. The accretion scheme is constructed in such a way as to explicitly conserve the linear momentum between the gas cells and the MBH particle. 

As a BH accretes, it will also release kinetic, thermal, and radiative energy in the form of jets or winds - so-called BH feedback. In this study, we have ignored this feedback mechanism and this caveat is discussed in detail in \S \ref{sec:caveats}. We will explore the impact of BH feedback on our results in a forthcoming study.

\subsection{Chemistry solver}
We model the chemistry using SGCHEM \citep{hartwig2015improved} as recently employed by \cite{wollenberg2020formation} and \cite{prole2022fragmentation}, which consists of 12 species, H, H\textsuperscript{+}, H\textsuperscript{-}, H\textsubscript{2}\textsuperscript{+} , H\textsubscript{2} , He, He\textsuperscript{+}, He\textsuperscript{++}, D, D\textsuperscript{+}, HD, and e\textsuperscript{-}. The primordial gas follows 45 reactions based on the chemical network described in \cite{clark2011formation} and updated by \cite{schauer2017formation}. The initial gas fractions are $2\times10^{-6}$ for H\textsubscript{2}, $10^{-4}$ for H\textsuperscript{+}, $2.6\times10^{-9}$ for D\textsuperscript{+}, and $2.6\times10^{-5}$ for D (the fraction is with respect to neutral hydrogen). The chemical network accounts for H\textsubscript{2} cooling, HD cooling, heating and cooling from gas, shocks, compression, and expansion of the gas, collisionally induced H\textsubscript{2} emission, ionisation, and recombination, and also heating from accretion luminosity. The adiabatic index of the gas is computed as a function of the chemical composition and temperature using the \arepo Harten-Lax-van Leer with contact and Alfv\'{e}n HLLD Riemann solver.
 
\section{Results}
\label{sec:results}

\begin{figure}
    \centering
    \includegraphics[width=\columnwidth]{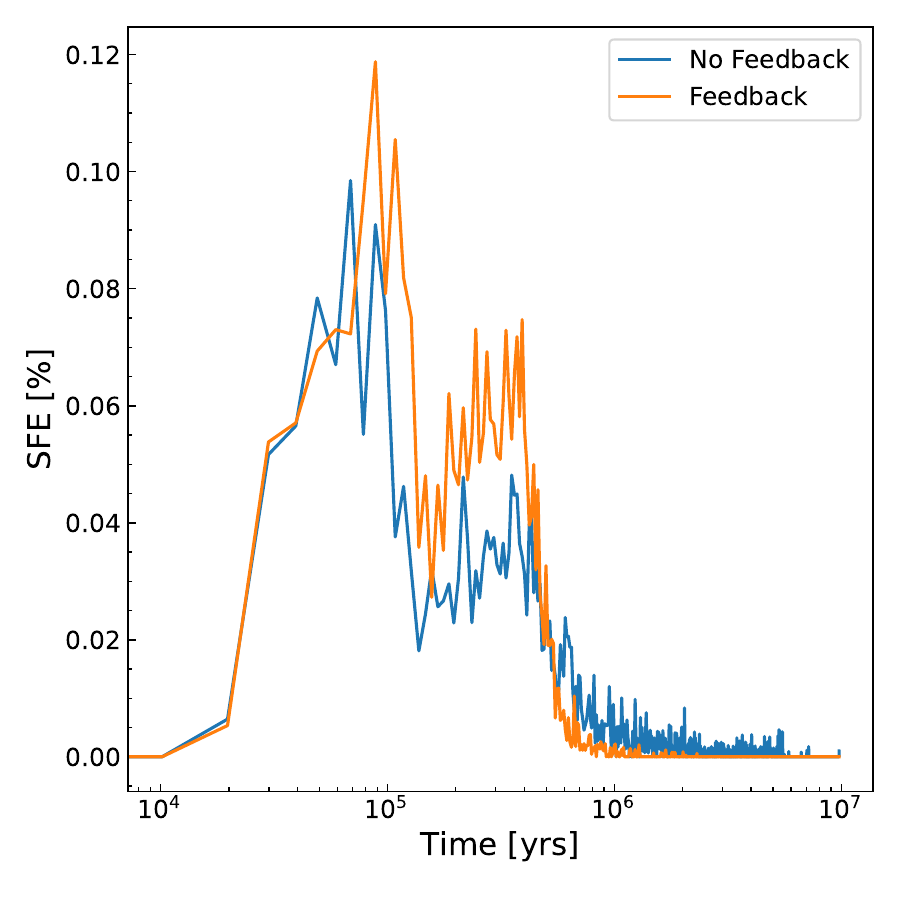}
    \caption{Star formation efficiency (SFE) for both the no-feedback (blue) and feedback (orange) simulations. We see a rise in SFE at the beginning, converting almost 0.1\% of the gas to stars. The SFE for feedback simulation is higher in the first $10^6$ yr overall. Nonetheless, the galaxy remains gas rich.}
    \label{fig:sfe}
\end{figure}

\begin{figure}
    \centering
    \includegraphics[width=\columnwidth]{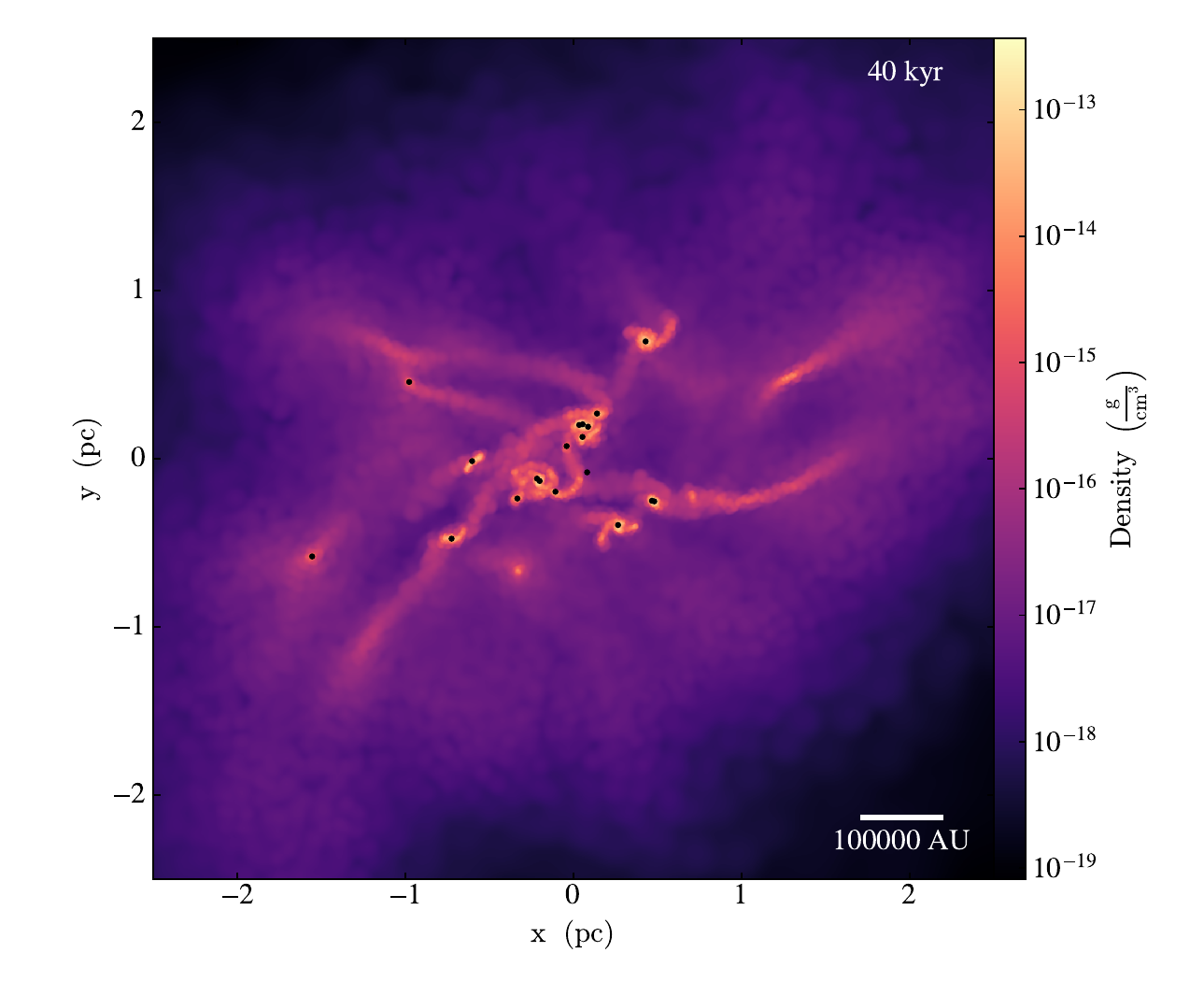}
    \caption{Sink formation (black dots) in a box of $5$ pc centred at the densest clump of the galaxy at T = $40$ kyr. The black dot represents PopIII particles. Even at this early stage in the evolution multiple PopIII stars have formed in the densest regions.}
    \label{fig:Sink_formation}
\end{figure}

 In Figs. \ref{fig:gas_projection} (face-on) and \ref{fig:gas_projection_edge} (edge-on), we show the state of the galaxy after allowing the galaxy to evolve at lower resolution for 100 Myr after the initial setup. We set the time at this point to be T = 0. From this point on, the simulation box is evolved at a resolution of $\Delta x = 10^{-2}$ pc. The galaxy clearly shows a turbulent planar structure. The top-right panel in both figures is a zoom-in image at the same time around the densest gas cell in the galaxy. The gas cloud shown here, along with multiple other clouds, forms spiral arms due to the rotation of the galaxy. This serves as the starting point for the next phase of the simulations, which we henceforth refer to in the cases where SNe feedback was and was not enabled as the `feedback' and `no-feedback' simulations, respectively. Note that all PopIII particles are converted into MBH particles following their stellar lifetimes in the no-feedback case, while their conversion or deletion (in the case of PISN) depends on their stellar mass in the feedback simulation. 
We show the star formation efficiency (SFE), defined as the mass converted to PopIII particles per unit gas mass, in Fig. \ref{fig:sfe} for both simulations. We see a sharp increase in SFE in the first $10^5$ yr reaching higher than 0.1\% for both simulations. Our ICs have multiple dense gas clumps, ideal for star formation bursts, resulting in this sharp increase. Once the available dense gas converts into PopIII particles, the SFE decreases. We also observe a small bump in SFE between 2-3$\times 10^5$ yr. At this point, the majority of PopIII particles have been converted to MBH particles, which accrete a lot of gas. The large scale gas movement bumps the SFE to 0.02-0.04 \%. Around the same time in the feedback simulation, a lot of PopIII particles undergo supernova. This feedback pushes and disturbs the gas, creating conditions for further star formation episodes. We see that the SFE increases to 0.06-0.07\% in the feedback simulation as a result. After $4 \times 10^5$ yr in both simulations, the SFE becomes smaller than $10^{-3} \%$. Overall, our SFEs remain fairly low (much less than 1\%).

In Fig. \ref{fig:Sink_formation}, we show a zoom-in image at $t \sim 40$ kyr for the no-feedback simulation. Annotated on top of the image are the locations of multiple sink particles - in this case PopIII particles.  In the image, we see the formation of several PopIII particles at the centre of fragmenting gas clouds. Recall that a PopIII particle has the same dynamical mass as its progenitor gas cell but is assigned a stellar mass by randomly sampling an IMF. These PopIII particles continue to evolve, with more PopIII particles forming over the course of the simulation. As this is for the no-feedback simulation, no SNe explosions have occurred or will occur.

\begin{figure}
    \centering
    \includegraphics[width=\columnwidth]{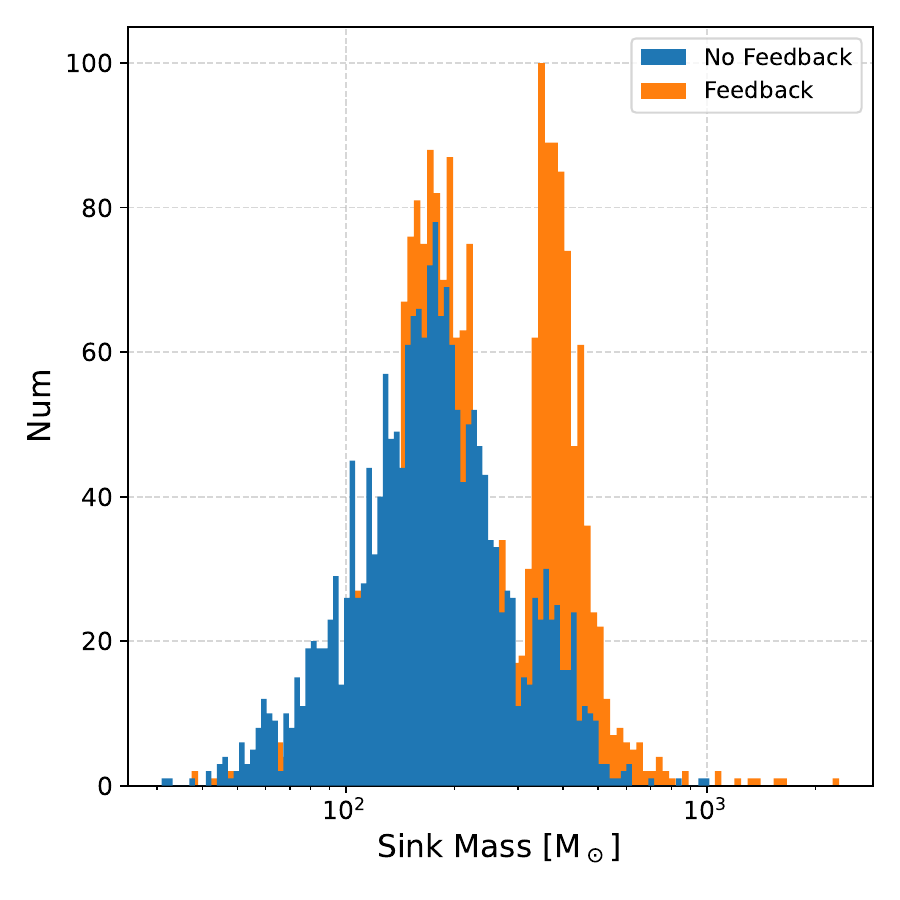}
    \caption{Initial PopIII particle masses for both the feedback (orange) and no-feedback (blue) simulations. Both initial distributions are broadly similar but the feedback case shows a distinctive bi-modal distribution. This bi-modality is caused by the higher temperature gas in the feedback case raising the Jeans mass and leading to higher (dynamical) PopIII particle masses for some of the objects.}
    \label{fig:Sink_population}
\end{figure}

\begin{figure*}
    \centerline{
    \includegraphics[width = \columnwidth]{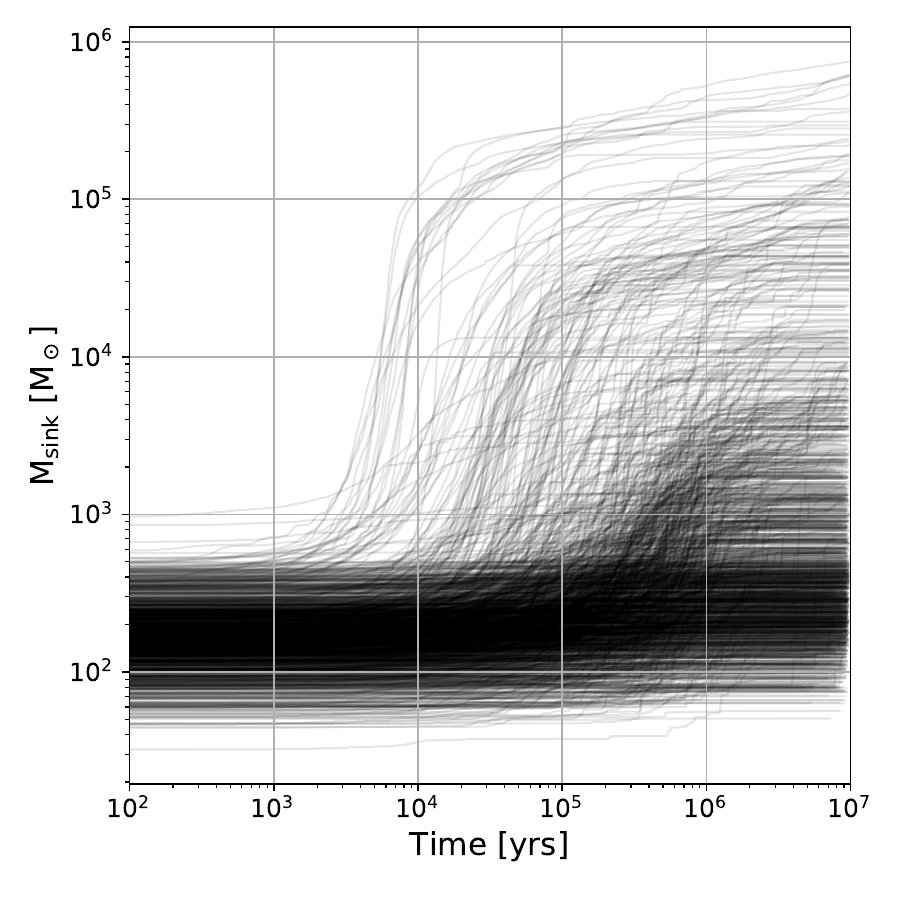}
    \includegraphics[width = \columnwidth]{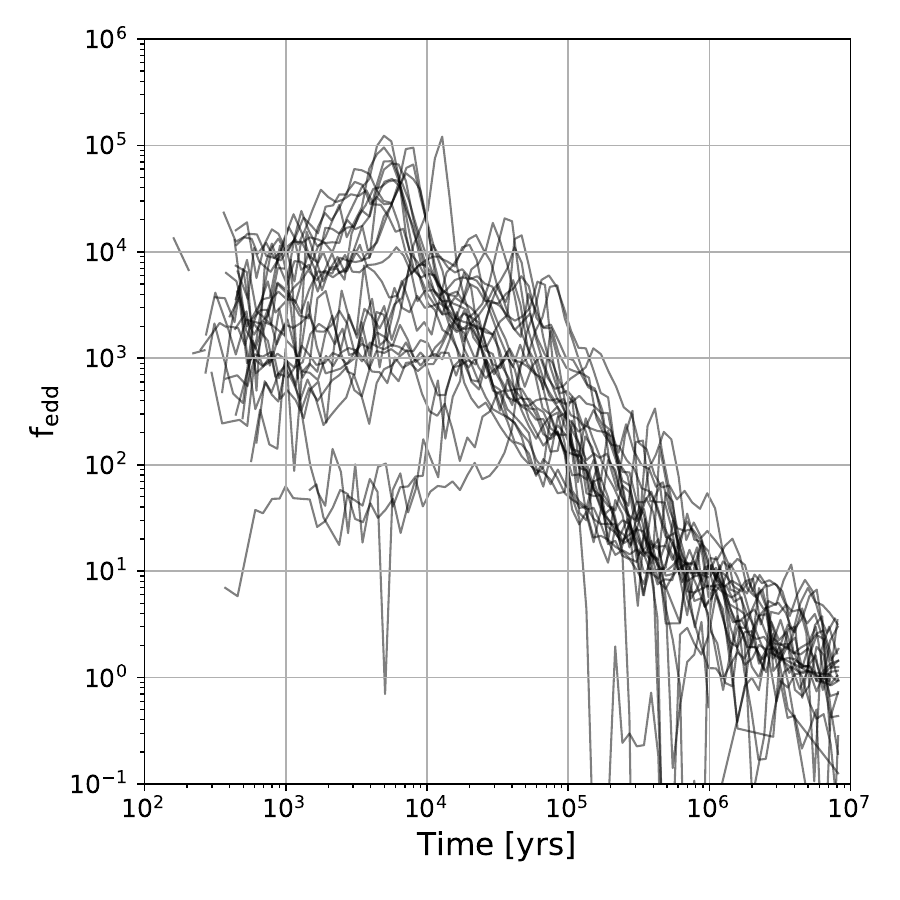}}
    \caption{Growth of MBH particle masses with no feedback. Left Panel: In the left hand panel we show the growth of the active sinks as a function of time. While most of the MBH particles experience negligible growth $\sim 38.5$ \% of all particles grow by more than 100 \%. Right Panel: The Eddington ratio for all of the MBH particles that reach a final mass of greater than $10^5$ \msolarc. The Eddington rates are initially hyper-Eddington and remain super-Eddington for up to 1 Myr before gas starvation slows down accretion. }
    \label{fig:Sink_growth_no_feedback}
\end{figure*}

In Fig. \ref{fig:Sink_population}, the initial PopIII particle masses are shown from both the no-feedback and feedback simulations. We form $2004$ and $2181$ PopIII particles in the no-feedback and feedback simulations, respectively, over the course of 10 Myr. 
A bimodal distribution is observed extending from $\sim 30$ to $\sim 10^3$ $\msun$ with two peaks around $\sim 200$ \msolar and $\sim 400$ $\msun$. The second peak PopIII particles are created from higher temperature gas cells, which has a higher Jeans density thereby impacting on the sink formation criteria via the Jeans density threshold. This means that higher temperature gas can lead to higher sink masses as expected. Consequently, in the feedback simulation, we get a higher number of massive PopIII particles and the bi-modal distribution is more prominent. In both cases, these are the dynamical masses of the PopIII particles. We now examine the results from the no-feedback and feedback simulations in turn. 

\subsection{No-feedback Simulation}

Once a PopIII particle reaches the end of its stellar lifetime, it is converted into a MBH particle, and it is able to accrete gas. We show the growth of MBH particles through accretion in Fig. \ref{fig:Sink_growth_no_feedback} for the no-feedback case. In the left panel we only show MBH particles that accrete more than $1 \msun$ over the course of the simulation - we henceforth refer to these as active sinks. The stellar phase of PopIII particles is not shown in this case, i.e. $t=0$ refers to the time at which the PopIII particle was converted into a MBH particle. Initially, accretion is suppressed as excess angular momentum must be shed before accretion can accelerate. However, as accretion occurs, the angular momentum of the gas slowly decreases, which increases the accretion rate, which in turn decreases the angular momentum. As a result, a rapid run-away accretion episode occurs, increasing the gas mass to an excess of $10^5$ $\msun$ for the most rapidly accreting MBH particles. 
A total of $1520$ MBH particles out of $2004$ accrete more than 1 solar mass (i.e. approximately 75\%). Of this total number of actively accreting seeds, 33 reach final masses in excess of $10^5\msun$. The most-massive MBH particle reaches a final mass of $\sim7.5 \times 10^5$ $\msun$. \\
\indent In the right hand panel, we show the Eddington ratio for the most rapidly accreting MBH particles - those that reach masses in excess of $10^5$ \msolar in less than 10 Myr. The accretion is initially in the hyper-Eddington regime, but as the surrounding gas mass is gradually depleted, the accretion rate falls to values closer to the Eddington rate as the initially plentiful supply of gas is consumed.

\begin{figure}
    \centering
    \includegraphics[width=\columnwidth]{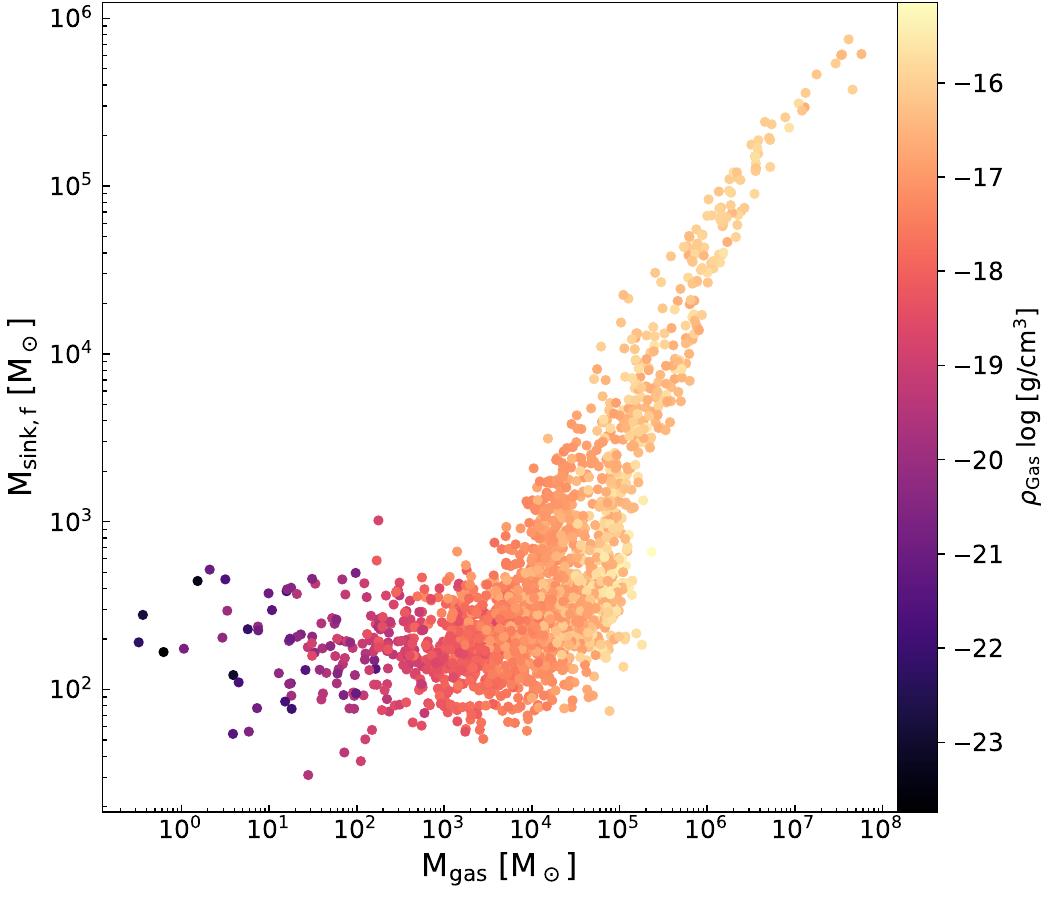}
    \caption{Final mass of the MBH particles v/s total gas mass within $r_{BH}$ of the MBH particles. The total gas mass is calculated from the entire simulation of $10^7$ yr. The colour bar shows the average density of the gas. The MBH particles that undergo runaway accretion capture $\sim 1\%$ of the total gas mass.}
    \label{fig:surrounding gas}
\end{figure}

\begin{figure*}
    \centering
    \includegraphics[width=\textwidth]{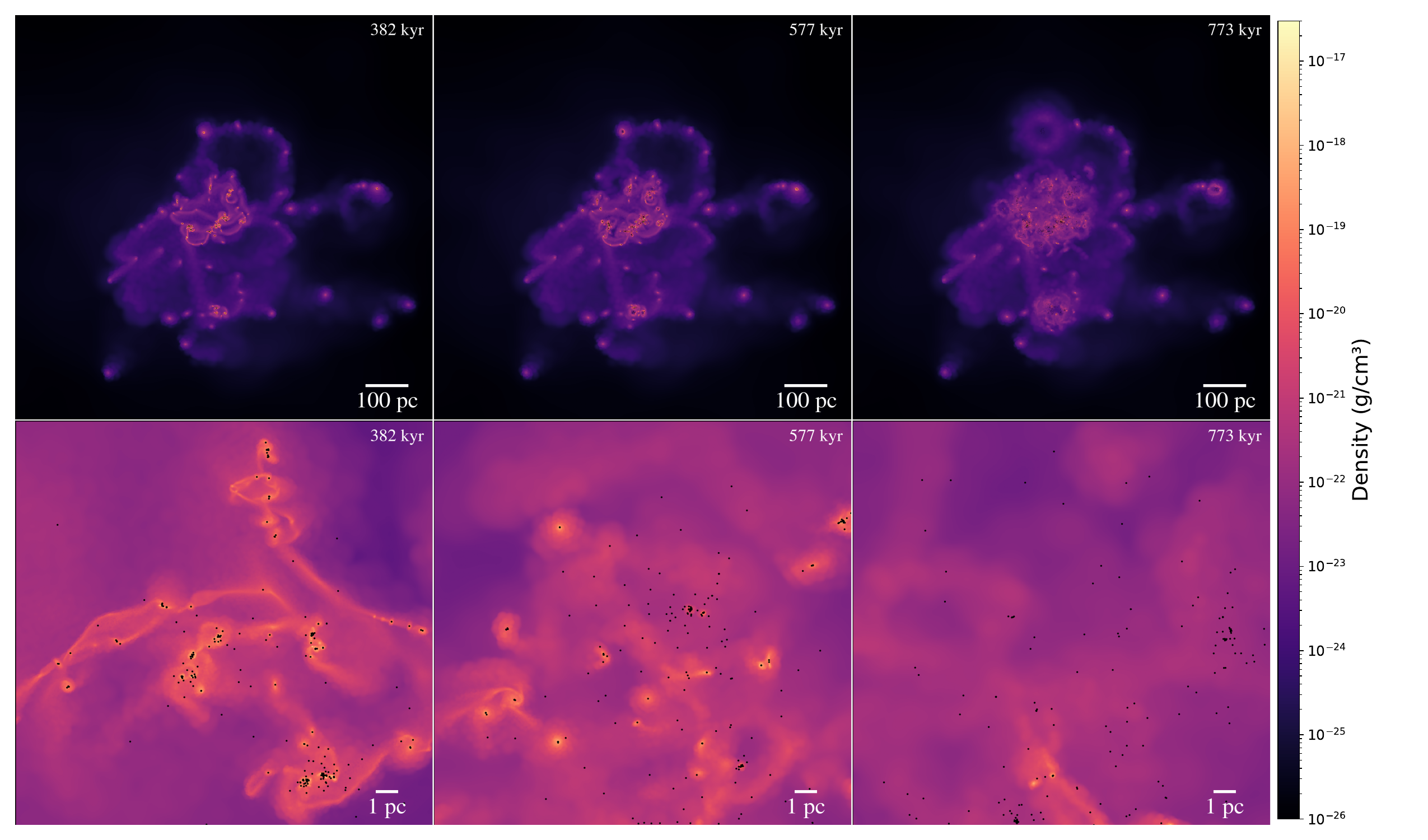}
    \caption{Time series projection plots for the supernova feedback simulation. All panels are centred at the densest clump in the IC file. The top panels show the entire galaxy and we zoom in to the densest clump in the bottom plots. The density projection is calculated the same way as Fig. \ref{fig:gas_projection}. The black dots represent the PopIII and MBH particles. We see on global scales, supernova feedback causes shells of gas to expand radially. On smaller scales, the clumps are completely disrupted.}
    \label{fig:supernova_feedback_visualize}
\end{figure*}

\begin{figure*}
    \centerline{
    \includegraphics[width = \columnwidth]{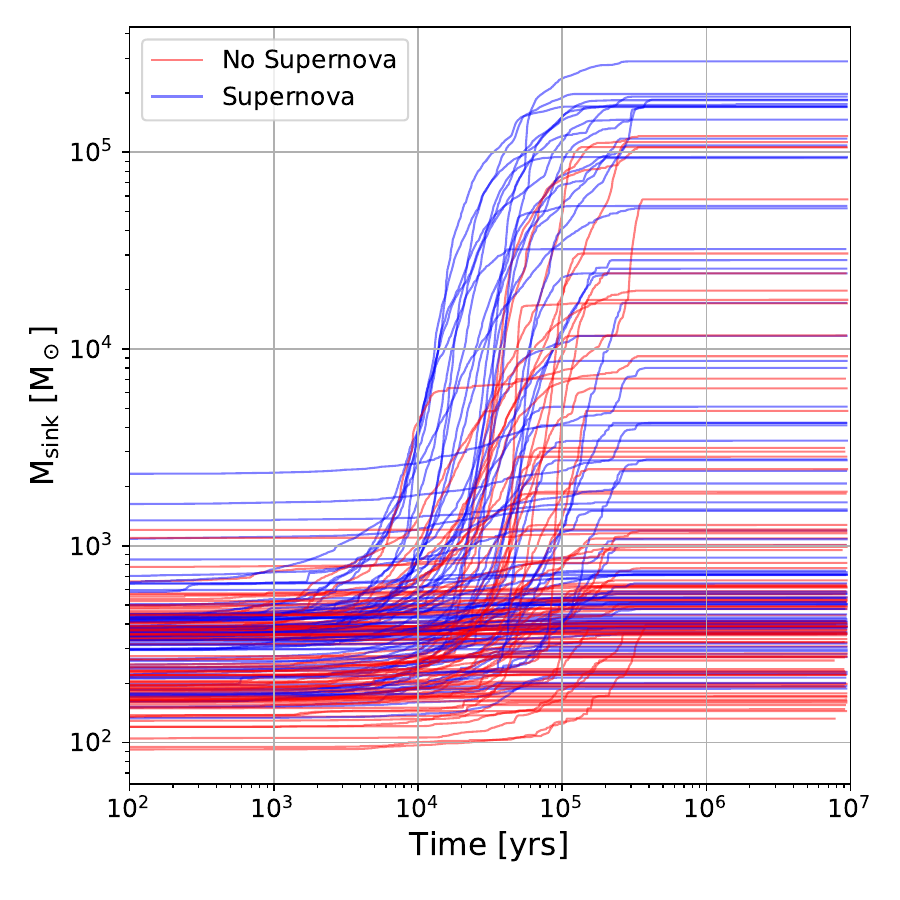}
    \includegraphics[width = \columnwidth]{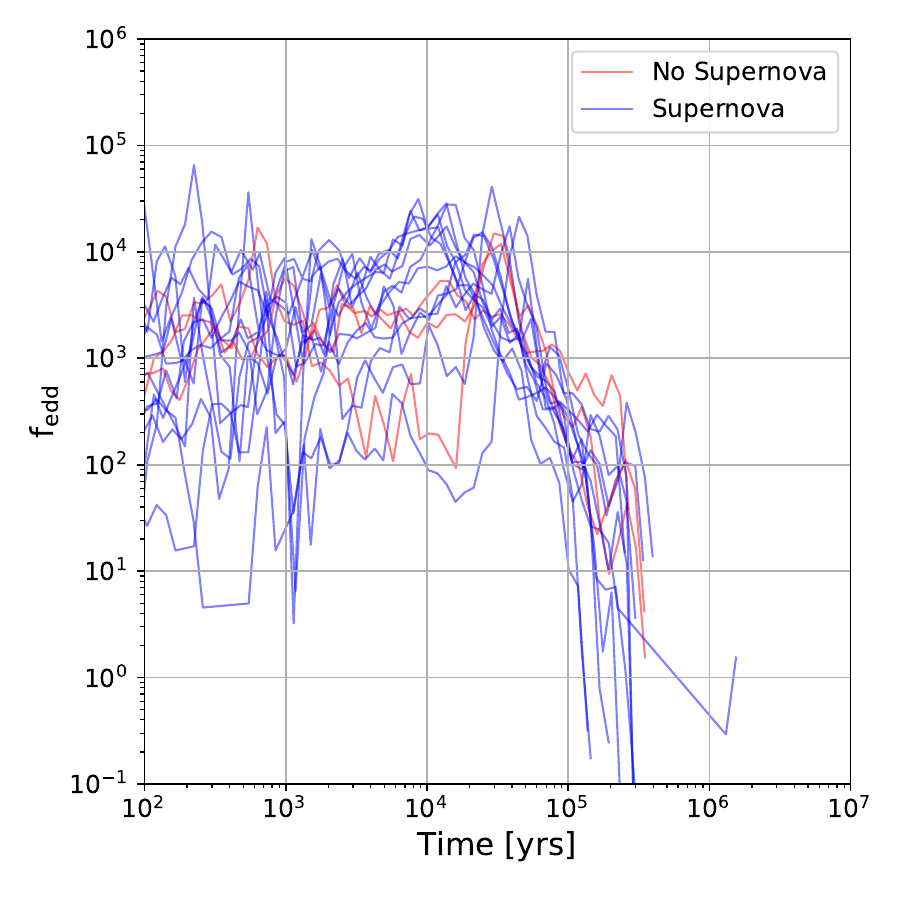}}
    \caption{Growth of MBH particle masses with feedback. Left Panel: In the left hand panel we show the growth of the active sinks as a function of time. Blue and red trajectories represent BH particles that formed through supernova and direct collapse, respectively. While most of the BH particles experience negligible growth $\sim 3.5$ \% of all particles grow by more than 100 \%. Right Panel: The Eddington ratio for all of the MBH particles that reach a final mass of greater than $10^5$ \msolarc. The Eddington rates are similarly hyper-Eddington at the beginning and remain super-Eddington for up to $\sim 2 \times 10^5$ yr before galaxy disruption by feedback halts accretion.}
    \label{fig:Sink_growth_feedback}
\end{figure*}

\begin{figure}
    \centering
    \includegraphics[width=\columnwidth]{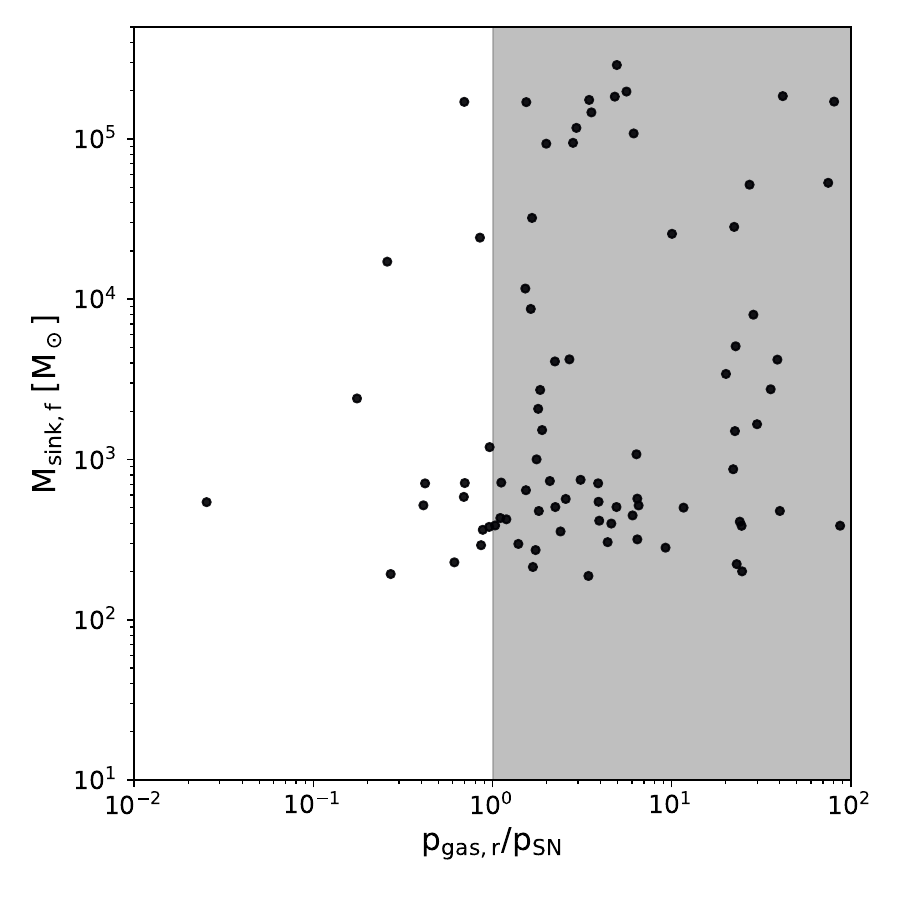}
    \caption{Final mass of the MBH particles as a function of the momentum ratio between inward gas momentum and supernova momentum. The radially inward momentum is calculated from the closest 1000 cells around the MBH particles. We see that for a majority of the MBH particles, the ratio is larger than 1, suggesting that supernova feedback was inefficient in pushing back the surrounding gas.}
    \label{fig:mom_injection}
\end{figure}

In Fig. \ref{fig:surrounding gas}, we show the final mass of the MBH particles as a function of total mass within its BHL radius ($r_{BH}$) throughout the simulation. The colourbar represents the average density of the gas surrounding the MBH particles. There are two features to note; first, we see that $1\%$ of the gas mass has been accreted onto the MBH particle on average. Secondly, the final mass is correlated with the average surrounding gas density, with the most rapidly growing MBH particles having the highest density gas surrounding them. 

To attain the largest masses, the progenitor PopIII particle must remain at the centre of the gas clumps they formed from. Due to the lack of feedback, there is no mechanism that pushes the gas back from the PopIII particle. When PopIII particles turn into MBH particles, the surrounding gas is therefore easily accreted; this will not always be the case in the feedback simulation (see \S \ref{sec:feedbackresults}). The presence of radiative feedback from PopIII particles would also push back the gas, which added with supernova feedback may decrease the number of growing BHs (see \S \ref{sec:caveats}).

There are 234 MBH particles surrounded by gas with density greater than $10^{-16}$ $\rm{g/ cm^{3}}$, which is approaching our maximum resolution. These gas cells should convert into a PopIII particle (due to the Jeans density criteria at our resolution), but because these gas cells lie within the accretion radius of another PopIII particle, they do not convert to PopIII particles. During the stellar lifetime of the PopIII particle, these gas cells accumulate more mass and their density increases. These cells then form the basis for rapid accretion following the collapse of the PopIII particle into a MBH particle. This build-up of mass surrounding the MBH is why we see initially rapid accretion (see again the right panel of Fig \ref{fig:Sink_growth_no_feedback}). Once these dense gas clouds have been accreted, which occurs on a timescale of approximately $10^5 - 10^6$ yr, then the growth slows dramatically with final masses between $10^4$ and $10^6$ \msolarc. We now examine the more realistic case where SNe feedback can influence and in many cases suppress rapid growth. 

\subsection{Feedback}
\label{sec:feedbackresults}
When the SNe feedback is turned on, PopIII particles may, depending on their stellar masses, undergo SNe explosions and inject momentum (Eq. \ref{eq:mom_inj}) into their surrounding gas. To see the impact of supernova feedback, we show face-on gas density projection plots in Fig. \ref{fig:supernova_feedback_visualize}. We show three instances of the simulation at $t=382, 577, 773$ kyr for the entire galaxy (top panel) and zoom in on the densest clump in the IC (bottom panel). The high resolution of our simulation allows us to see the impact of individual SNe explosions. When a SNe explosion occurs, the gas surrounding the MBH particles expands spherically while transferring the momentum in radially outward directions. This can be seen in the top-right panel at $t=773$ kyr where a large supernova occurs at the top side of the galaxy. Due to multiple supernovae occurring in a short time period, the gas is injected with a lot of energy, which completely disrupts the galaxy. This can be seen in the bottom panels, where the dense clumps and filaments are completely disrupted by the supernovae. Recall that not all PopIII particles undergo a SNe explosion - see \S \ref{sec:supernova} for details. Out of 2181 PopIII particles, 99 experienced PISN and were deleted from the simulation.

In the left panel of Fig. \ref{fig:Sink_growth_feedback}, the growth of MBH particles following their stellar phase is shown when the SNe feedback is turned on. The particles that undergo a SNe explosion (and subsequently transition into a MBH particle) are shown in blue. The particles that directly collapse into a MBH are shown in red. At first glance, we can tell that just by turning on SNe feedback, the number of MBH particles that accrete over $1\msun$ has decreased by almost an order of magnitude compared to the no-feedback case. Only $184$ MBH particles grow by more than $1\msun$ - compared to 1520 in the no-feedback case. However, $16$ of these particles were still able to grow beyond $10^5\msun$. Furthermore, feedback delays growth as MBH particles reach $10^5$ $\msun$ in $10^5$ yr, which is an order of magnitude longer than in the no-feedback case. \\
\indent Out of the 184 MBH particles that grow by more than a solar mass, 101 form through the direct collapse black hole (DCBH) channel, where there is no SNe feedback and the PopIII particle directly collapses into a MBH. As a consequence, the surrounding gas does not experience any push back. These MBH particles are similar to the no-feedback simulation, and their growth trajectories are shown in red in the left panel of Fig. \ref{fig:Sink_growth_feedback}. A universal trend for the growth is the appearance of flat line after a few $10^5$ yr because the galaxy is completely disrupted by the SNe and MBH particles find efficient accretion more difficult to attain. \\
\indent In the right panel of Fig. \ref{fig:Sink_growth_feedback}, we show the Eddington ratio for each of the active sinks as a function of time. Similarly to the no-feedback case, MBH particles initially grow extremely rapidly with Eddington ratios between $10^2$ and $10^4$. We see that for MBH particles that undergo SNe, the $\rm{f_{edd}}$ begins at an earlier time compared to the no-feedback case. Furthermore, SNe feedback decreases $\rm{f_{edd}}$ by an order of magnitude compared to the no-feedback case. Interestingly, MBH particles that undergo SNe accrete more rapidly than those formed via the DCBH channel, i.e. the rapid growth is dominated by blue lines. \\
\indent To understand why the exploding particles grow despite SNe feedback, we show in Fig. \ref{fig:mom_injection} the final mass of the MBH particles as a function of the ratio between the gas momentum in the radially inward direction towards the MBH particle and the momentum injected by the SNe. The gas momentum is calculated for the closest $1000$ gas cells to the progenitor PopIII particles just before the SNe explosions. The grey region is where the radially inward momentum is greater than the SNe feedback. We see that for a majority of MBH particles that grow, the radially inward momentum is larger than the SNe momentum, so feedback is inefficient in pushing back the gas. Furthermore, because the gas had expanded and re collapsed, its angular momentum had decreased. However, this does not necessarily lead to increased accretion rates as the newly ionised gas is hotter and takes more time to collapse. Finally, in Fig. \ref{fig:mom_injection}, we also notice a few points outside the grey area, suggesting that these MBH particles grew despite SNe successfully pushing back the surrounding gas. We find that for these particles, the shock waves produced by SNe originating from other PopIII particles had pushed gas in their direction, allowing them to quickly capture and accrete from it.

\section{Discussion}
\label{sec:discussion}
\noindent The summary result from this study is that we observe multiple cases of rapid, hyper-Eddington, accretion onto light seeds buried in a  dense gas-rich environment. One reason for the rapid growth we observe is that the PopIII particles do not easily decouple from the gas clumps in which they are formed. The size of the gas clump depends on the minimum resolution of our simulations - this is a limitation of our simulations (and many others). Increasing the resolution would cause further fragmentation and formation of even smaller gas clumps. This potentially could decrease the amount of gas a MBH particle would accrete in the runaway accretion episode. A higher resolution realisation may also result in the formation of more MBH particles, dividing the gas available for accretion by an individual MBH particle. \\
\indent In any case and independent of the caveats regarding resolution, the results we show here are important given the outstanding question of whether light seeds or heavy seeds are the dominant pathways to massive black holes in galaxy centres \citep[e.g.][]{regan2024massive}. It is however, not the first time that the growth of light seeds has been demonstrated in numerical models \cite[e.g.][]{gordon2024hungry, shi2023hyper, smith2018growth, Lupi_2016}. We now briefly review and discuss the existing literature and results and put the results from this study in context. 


Recently, \citet{gordon2024hungry} studied the growth of light seeds in mini-halos at a higher resolution of $\Delta x \approx 10^{-4}$ pc. Using a similar sink formation and accretion prescription to us within the \enzo code, the MBH particles in their simulation similarly start accreting in high density regions, able to grow to approximately 2000-3000 $\msun$ within 1 Myr. However, accretion halts because a thick disk forms around the MBH particles, which increases the gas's angular momentum. Their MBH particles fail to grow beyond a few times $10^3$ $\msun$ for two reasons: i) their higher resolution ($100 \times$ our resolution) causes fragmentation to form smaller gas clouds, and ii) this higher resolution also allows the formation of an accretion disk around the MBH particles which we do not see. The MBH particles are then unable to accrete gas and instead the gas accumulates on an accretion disk. 

An even higher resolution study was performed by \citet{prole2022fragmentation} with a resolution of $10^{-7}$ pc ($10^5 \ \times$ our resolution). Their gas clumps were so small that accretion was of the order of $10-100$ $\msun$. This study showed that metal-free gas continues to fragment down to sub-au scales where the collapse becomes adiabatic. As these simulations were only followed for short timescales ($\sim100$ yr), it is unknown whether the clumps formed at our resolution share the gas mass between a growing number of unresolved objects, or if long term accretion and violent mergers within the clump result in the formation massive objects with masses comparable to that of our sink particles. 

\begin{figure}
    \centering
    \includegraphics[width=\columnwidth]{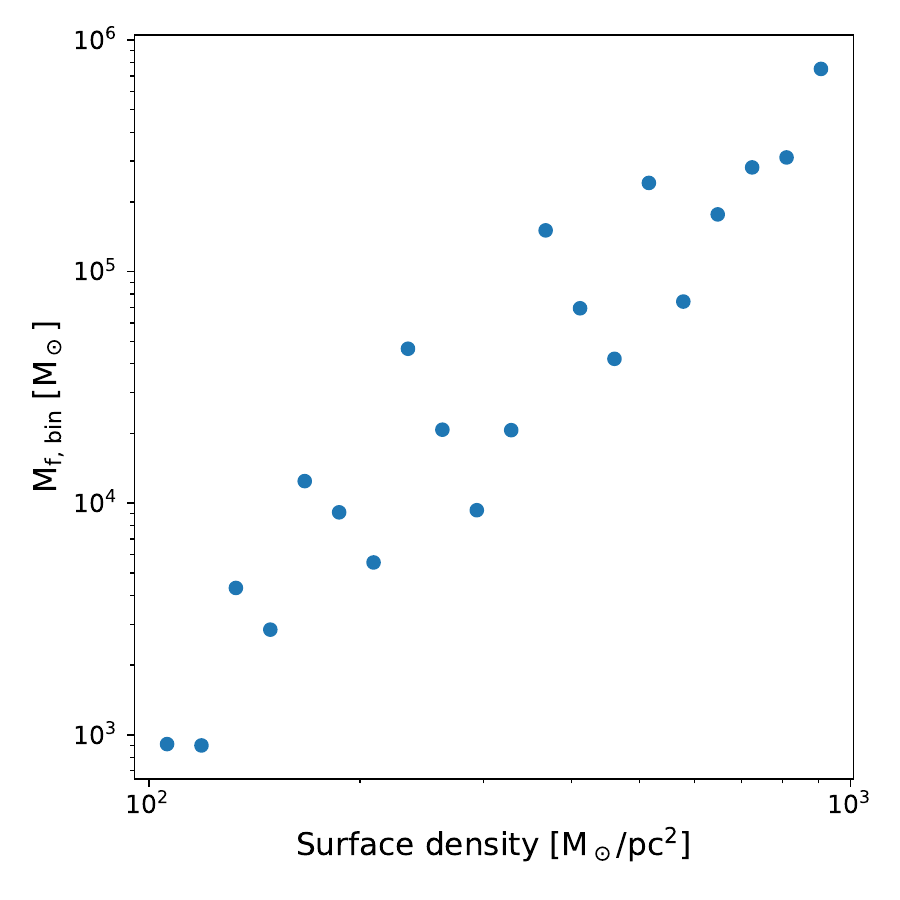}
    \caption{Largest MBH particles formed as a function of surface density. We bin the surface density in logarithmic space into 20 bins and select the largest MBH particle for that bin. The surface density is calculated for the central 100 pc of the galaxy at the moment when the PopIII particle is converted to a MBH particle. }
    \label{fig:surface_density}
\end{figure}

The most relevant and directly comparable work to ours is that of \cite{shi2023hyper}, who studied the growth of BH particles in large star forming molecular clouds and found hyper-Eddington growth. They found that sustained hyper-Eddington accretion occurs when the following conditions are satisfied: i) the BH particles find themselves in a dense region with gas density $\sim 100$ times that of the mean density, ii) low relative velocities between the BHs particles and the dense gas clumps (The BH particles are captured by the gas clumps). Both of these conditions are easily satisfied by the MBH particles in our simulations, which do not decouple from their formation gas clump. They also find that the surface density of the entire cloud determines the maximum mass a BH particle can grow to. In Fig. \ref{fig:surface_density}, we show the maximum mass of MBH particles formed as a function of surface density of the galaxy at that time. We calculate the surface density within the central $100$ pc of the galaxy. Similarly to \cite{shi2023hyper}, we find that the final mass of the MBH particle is correlated with the surface density of the galaxy. Additionally, we observe that the final mass of the MBH particles is independent of the initial MBH particle masses (mass immediately after the stellar lifetime ends). Another similarity arises from the dual role of feedback on accretion. While SNe feedback suppresses accretion overall, it also creates regions of colliding shock waves that prove excellent for gas capture by BH particles - again consistent with the results from \cite{shi2023hyper}.

Another study conducted by \citet{Lupi_2016} found light seeds growth in the nuclear region of high-redshift massive spiral galaxies. They performed Eulerain simulations using \texttt{Ramses} \citep{teyssier2002cosmological} and Lagrangian simulations using \texttt{Gizmo} \citep{hopkins2014gizmo} at a resolution range of $0.4$ pc to $0.02$ pc (comparable if a little below ours). They show that for BH particles to grow, radiatively inefficient accretion is necessary \citep[e.g.][]{Madau_2014}. However, this is not a sufficient condition - they additionally find that the BH particle must bind to the dense gas, consistent with our results. They additionally found that the gas must be dense enough to survive the impact of feedback, again consistent with our results. As our simulations do not have radiative BH feedback, we satisfy all of the conditions described. \cite{Lupi_2016} show extremely rapid growth of light seeds up to masses of approximately $10^{3-4}$ \msolar depending on the different simulation setups and resolution.

While the above (full hydrodynamic) studies all show rapid growth of light seeds is possible - there are clearly differences in the subgrid models employed and more importantly in the resolution in each case. Higher resolution simulations show lower overall accretion rates (e.g. \citealt{gordon2024hungry} and \citealt{prole2022fragmentation}). It is therefore a caveat of our work that a higher resolution study may indeed lead to higher levels of fragmentation and less growth for our MBH particles. However, lower resolution simulations also show conflicting results and lower levels of growth.  \cite{smith2018growth} when analysing the \renaissance simulation suite \citep{xu2013population, xu2014heating, chen2014scaling, ahn2015spatially, o2015probing, xu2016galaxy, xu2016late} observes no significant growth in any of their more than 20,000 BH particles spread across a range of high redshift galaxies. The reasons for this are likely to be multiple. While their resolution is relatively low (the densities they work with are between $10^{-25}$ to $10^{-23}$ $\rm{g/cm^3}$, almost 7 orders of magnitude lower than ours), this could be expected to increase rather than suppress growth. \\

\indent However, even with a boost factor of $100-200$ in accretion rates, they do not get as high accretion rate as we see. Furthermore, \cite{smith2018growth} observed that their BH particles did not encounter high densities regions within their cosmological halos. One of the reasons for this is due to the radiation feedback from their progenitor stars, a feature we do not model. The main difference is likely to boil down to the environments considered. The surface mass densities reached in our work are considerably higher than those seen in the \renaissance suite studied by \cite{smith2018growth}, and hence it may be that they either did not resolve or encounter the correct environment. Hence, while their study showed that growth is not obvious in a cosmology context, we show that given the correct conditions, rapid accretion becomes possible and BHs can grow at super (hyper)-Eddington rates. We also emphasise that our study (as many of the studies which show light seed growth) is highly idealised. A more self-consistent study is needed to allow a more robust comparison.  

\section{Caveats and Future Work}
\label{sec:caveats}
Our study, given its idealised setup, has a number of caveats, which we now discuss. 
The first caveat of our study is the short timescale on which we run the simulations ($\sim 10$ Myr). As a consequence, we shorten our stellar lifetimes by a factor of 100 so that the remnant BH particles form early. This leads to the majority of PopIII particles having lifetimes shorter than $10^5$ yr. This in turn means that PopIII stars have very little time to decouple from their formation gas clumps. Is it possible that longer (more realistic) lifetimes may result in the PopIII stars decoupling from their formation cloud and subsequently the remnant BH being born into lower density environments. This would lead to less BHs reaching $10^5 \msun$. Additionally, in the feedback simulations, because of shorter stellar lifetimes, we have multiple supernovas going off within a short time, leading to complete disruption of the galaxy. This, of course, has the opposite effect and potentially suppresses growth.

Another caveat of our work is the absence of radiative feedback from PopIII particles. Including radiative feedback from stellar populations would bring our work closer to that of \renaissance \citep[e.g.][]{smith2018growth}. We plan to do this in a forthcoming study. The additional realism of progenitor radiative feedback may contribute to decreasing the accretion rate because the gas clumps may be disrupted by this feedback even before the SNe feedback sets in. However, this is not a certainty. High resolution simulations of the effects of radiative feedback from PopIII stars have shown formation of small H II regions around PopIII stars, with the radiation remaining trapped inside dense gas clumps \citep{jaura2022trapping}.

In addition to neglecting feedback from the stellar populations, we neglect feedback from the accreting black holes. This again would work to suppress accretion and is a limitation of our work here (and may set an upper limit to our results).

In future work, we will explore the growth of light seed in a cosmological context, including the impact of radiative feedback from PopIII particles and from the accreting BHs. By expanding our dataset to a cosmological setting, we plan to explore the conditions required for light seed growth in a more self-consistent and less idealised fashion.

\section{Conclusions}
\label{sec:conclusion}
\noindent In this work, we investigated whether the conditions necessary for light seed growth occur within gas-rich galaxies at high redshift. Using simulations performed with the \arepo code \citep{Springel_2010, Pakmor_2016}, we modelled the creation and accretion behaviour of sink particles, which represent stars during their lifetimes and later evolve into BHs.

The initial galaxy, with a mass of approximately $10^8 \msun$, was generated using GalIC \citep{yurin2014iterative}, allowing for a stable and self-consistent initial condition. Moreover, the galaxy setup we employ models the chaotic, gas rich and relatively compact galaxies expected at high-z. Two realisations of this galaxy were explored: one where SNe feedback was disabled, and another where SNe feedback was enabled. These two configurations allowed us to assess the impact of SNe feedback on the environments and subsequent accretion that could support light seed formation and growth. In this work, we have neglected radiative feedback from PopIII stars. We also employed a stellar lifetime shortening to properly capture the BH phase of a sink particle. Both of these caveats may alter our results and reduce the number of growing BHs or their final masses, we will investigate the impact of both PopIII radiative feedback and BH feedback in future works. The following are our conclusions.

\begin{itemize}
    \item Our main conclusion is that stellar mass BHs (light seeds) can grow extremely efficiently and rapidly inside gas rich environments. Moreover, we find that approximately 1\% of our light seeds can accrete to masses in excess of $10^4$ \msolar inside 10 Myr. 
    \item In the realisation with no-feedback, we formed $2004$ MBH particles among which $1520$ were able to accrete more than $1$ $\msun$. $33$ MBH particles grew larger than $10^5$ $\msun$. The fastest growing MBH particles grew larger than $10^5$ $\msun$ within just $\sim 10^4$ yr. Such rapid accretion is observed because these particles do not decouple from the gas clump they were formed from. In the absence of feedback, they rapidly accrete the entire gas clump. These MBH particles have initial Eddington factors between $10^2-10^4$, and the BHs grow to $\sim 10^5$.

    \item The realisation with feedback formed $2082$ MBH particles (about 100 PopIII particles underwent PISN and were deleted) and $184$ of them accreted more than $1 \msun$. The addition of feedback decreased the number of growing MBH particles by an order of magnitude. However, $16$ MBH particles were still able to grow beyond $10^5$ $\msun$. Some of these particles formed through the direct collapse channel, so they experienced no feedback. For the remainder, we find that the SNe feedback was ineffective in pushing back the surrounding gas, causing it to recollapse on the MBH particles. This expansion and recollapse delay growth to $10^5 \msun$ but does not stop it.

    \item The growth of some MBH particles was also enhanced by SNe feedback. Shocks push gas towards MBH particles while colliding shocks from several SNe also created regions ideal for runaway accretion to occur.
     
\end{itemize}
In summary, we find that even in the presence of SNe feedback, light seed growth is possible for approximately 1\% of BHs in dense, gas rich environments at high redshift. The total halo mass of our realisation is approximately $10^8$ \msolar and these galaxies are typical of proto-galaxies expected at high-z. Given the right external conditions, such galaxies are likely ideal incubators for growing the first MBHs starting from a light seed. In a forthcoming study we will place these results in a cosmological context and compare heavy seed number densities directly against light seed number densities with the goal of determining the breakdown of which environments will fulfil the requirements of light seed growth vs. heavy seed formation and growth. 


\section*{Acknowledgements}
\noindent The authors thank the referee for providing constructive feedback that greatly improved the manuscript.  JR acknowledges support from the Royal Society and Science Foundation Ireland under grant number 
 URF\textbackslash R1\textbackslash 191132. DM, JR and LP acknowledge support from the Irish Research Council Laureate programme under grant number IRCLA/2022/1165.
 \ \
The simulations were performed on the Luxembourg national supercomputer MeluXina and the Czech Republic EuroHPC machine Karolina.
The authors gratefully acknowledge the LuxProvide teams for their expert support.
\ \ 
The authors wish to acknowledge the Irish Centre for High-End Computing (ICHEC) for the provision of computational facilities and support.
\ \
The authors acknowledge the EuroHPC Joint Undertaking for awarding this project access to the EuroHPC supercomputer Karolina, hosted by IT4Innovations through a EuroHPC Regular Access call (EHPC-REG-2023R03-103)

\newpage

\bibliographystyle{mnras}
\bibliography{Bib}

\end{document}